\documentstyle[12pt]{article}
\input epsf.sty
\def\ZZZ{{\hbox{ Z\kern-1.6mm Z}}}

\newcommand{\eps}{\epsilon}

\newcommand{\LL}{{\cal L}}

\newcommand{\NN}{{\cal N}}

\newcommand{\be}{\begin{equation}}
\newcommand{\ee}{\end{equation}}
\newcommand{\ben}{\begin{eqnarray}\displaystyle}
\newcommand{\een}{\end{eqnarray}}
\newcommand{\refb}[1]{(\ref{#1})}
\newcommand{\p}{\partial}
\newcommand{\sectiono}[1]{\section{#1}\setcounter{equation}{0}}

\def\one{{\hbox{ 1\kern-.8mm l}}}
\def\zero{{\hbox{ 0\kern-1.5mm 0}}}

\begin{document}
{}~
{}~
\hfill\vbox{\hbox{hep-th/0506177}
}\break

\vskip .6cm

{\baselineskip20pt
\begin{center}
{\Large \bf
Black Hole Entropy Function and the Attractor Mechanism 
in Higher Derivative Gravity
} 

\end{center} }

\vskip .6cm
\medskip

\vspace*{4.0ex}

\centerline{\large \rm
Ashoke Sen}

\vspace*{4.0ex}

\centerline{\large \it Harish-Chandra Research Institute}

\centerline{\large \it  Chhatnag Road, Jhusi,
Allahabad 211019, INDIA}

\vspace*{1.0ex}

\centerline{\it and}

\vspace*{1.0ex}

\centerline{\large \it Center for Theoretical Physics, MIT, 
Cambridge,  MA 02139, USA}

\vspace*{1.0ex}

\centerline{E-mail: ashoke.sen@cern.ch,
sen@mri.ernet.in}

\vspace*{5.0ex}

\centerline{\bf Abstract} \bigskip

We study extremal black hole solutions in $D$ dimensions with near horizon
geometry $AdS_2\times S^{D-2}$ in higher derivative gravity coupled to
other scalar, vector and anti-symmetric tensor fields.  We define an
entropy function by integrating the Lagrangian density over $S^{D-2}$ for
a general $AdS_2\times S^{D-2}$ background, taking the Legendre transform
of the resulting function with respect to the parameters labelling the
electric fields, and multiplying the result by a factor of $2\pi$. We show
that the values of the scalar fields at the horizon as well as the sizes of
$AdS_2$ and $S^{D-2}$ are determined by extremizing this entropy function
with respect to the corresponding parameters, and the entropy of the black
hole is given by the value of the entropy function at this extremum. Our
analysis relies on the analysis of the equations of motion and does not
directly make use of supersymmetry or specific structure of the higher
derivative terms.

\vfill \eject

\baselineskip=18pt

\tableofcontents

\sectiono{Introduction and Summary} \label{s1}

Analysis of supersymmetric black holes in string theory have led to many
new insights into the classical and quantum aspects of black holes. In
particular a rich structure has emerged in the context of half-BPS black
holes in $\NN=2$ supersymmetric string theories in four dimensions. One of
the important features of these black holes is the attractor mechanism
\cite{9508072,9602111,9602136} by which the values of the scalar fields at
the horizon are determined only by the charges carried by the black hole
and are independent of the asymptotic values of the scalar fields.  The
entropy of these black holes agrees with the microscopic counting of the
states of the brane system they describe, not only in the supergravity
approximation, but also after the inclusion of higher derivative
corrections to the generalized
prepotential\cite{9711053,9602060,9603191,9801081,9812082,9904005,
9906094,9910179,0007195,0009234,0012232}. More recently it has been shown
that the Legendre transform of the black hole entropy with respect to the
electric charges is directly related to the generalized prepotential, and
this has led to a new conjectured relation between the black hole entropy
and topological string partition
function\cite{0405146,0412139,0502211,0504221}.  Finally, applying the
results for these black holes to the special case of black holes in
heterotic string theory with purely electric charges, one finds agreement
between black hole entropy and the degeneracy of elementary string
states\cite{0409148,0410076,0411255,
0411272,0501014,0502126,0502157,0504005,0505122} even though the black
hole entropy vanishes in the supergravity
approximation\cite{9504147,9506200,9712150}.

All of these results have been derived by making heavy use of
supersymmetry. In particular while taking into account the effect of
higher derivative terms one includes in the string theory effective action
only a special class of terms which can be computed using the partition
function of topological string theory\cite{9302103,9307158,9309140}.  
These corrections are controlled by a special function known as the
generalized prepotential\cite{9812082,0007195,0009234}.  While these
constitute an important set of terms in the string theory effective acion,
they are by no means the only terms, and at present there is no
understanding of why these terms should play a special role in the study
of black holes. In fact there are counterexamples, involving elementary
string states in type II string theory, for which the corrections to the
generalized prepotential are not enough to produce the desired result for
the black hole entropy\cite{0502126}.  Thus it seems important to study
the role of the complete set of higher derivative terms on the near
horizon geometry of the black hole.

In this paper we study the effect of higher derivative terms on the
entropy of extremal black holes in $D$ dimensions following the general
formalism developed in \cite{9307038,9312023,9403028,9502009}.  We do not
make use of supersymmetry directly, but define extremal black holes to be
those objects whose near horizon geometry is given by $AdS_2\times
S^{D-2}$.\footnote{Eventually supersymmetry may play a role in
establishing the existence of a solution that interpolates between the
near horizon $AdS_2\times S^{D-2}$ geometry and the asymptotic Minkowski
space-time.} 
We also define the entropy of the extremal black hole to be the
extremal limit of the entropy of a non-extremal black hole so that
we can use the general formula for the entropy given in
\cite{9307038,9312023,9403028,9502009} even though strictly extremal
black holes do not have a bifurcate horizon.
Our main results may be summarized as follows.

\begin{enumerate}

\item Let $S_{BH}(\vec q, \vec p)$ denote the entropy of a D-dimensional
extremal black hole, with near horizon geometry $AdS_2\times S^{D-2}$, as
a function of electric charges $\{q_i\}$ associated with one form gauge
fields and magnetic charges $\{p_a\}$ associated with $(D-3)$ form gauge
fields.  We choose a coordinate system in which the $AdS_2$ part of the
metric is proportional to $-r^2 dt^2 + dr^2/r^2$. Then the Legendre
transform of $S_{BH}(\vec q, \vec p)/2\pi$ with respect to the variables
$q_i$ is equal to the the integral of the Lagrangian density over the
$(D-2)$ dimensional sphere $S^{D-2}$ enclosing the black hole. The
variable conjugate to $q_i$ represents the radial electric field $e_i$ at the
horizon associated with the $i$-th gauge field.

\item Consider a general $AdS_2\times S^{D-2}$ background parametrized by
the sizes of $AdS_2$ and $S^{D-2}$, the electric and magnetic fields and
the values of various scalar fields. We define an entropy function by
integrating the Lagrangian density evaluated for this background over
$S^{D-2}$, taking the Legendre transform of this integral with respect to
the parameters $e_i$ labelling the electric fields and multiplying the
result by $2\pi$. The result is a function of the values $u_s$ of the
scalar fields, the sizes $v_1$ and $v_2$ of $AdS_2$ and $S^{D-2}$, the
electric charges $q_i$ conjugate to the variables $e_i$, and the magnetic
charges $p_a$ labelling the background magnetic fields. We show that for
given $\vec q$ and $\vec p$, the values $u_s$ of the scalar fields as well
as the sizes $v_1$ and $v_2$ of $AdS_2$ and $S^{D-2}$ are determined by
extremizing the entropy function with respect to the variables $u_i$,
$v_1$ and $v_2$. Furthermore the entropy itself is given by the value of
the entropy function at the horizon.

\item For extremal black hole solutions without Ramond-Ramond (RR) charges
in tree level string theory the Lagrangian density at the horizon vanishes
due to the dilaton field equation. In this case the entropy of the black
hole is given simply by $2\pi$ times the product of the electric field at
the horizon and the electric charge of the black hole.

\end{enumerate}

These results rely on the assumption that the Lagrangian density can be
expressed in terms of gauge invariant field strengths and does not involve
the gauge fields explicitly. Thus if Chern-Simons terms are present we
either need to remove them by going to the dual field variables, or if
that is not possible, consider black hole solutions which are not affected
by these Chern-Simons terms.

\sectiono{Entropy of Extremal Black Holes} \label{s2}

We begin by considering
a four dimensional theory of gravity coupled
to a set of  abelian gauge fields $A_\mu^{(i)}$
and 
neutral scalar fields $\{\phi_s\}$.  Suppose $\sqrt{-\det g}\,
\LL$ is the lagrangian density, 
expressed as a function of the metric $g_{\mu\nu}$, the scalar fields 
$\{\phi_s\}$,
the gauge field strengths $F^{(i)}_{\mu\nu}$, 
and covariant derivatives of these fields.
We consider a spherically symmetric extremal black hole solution
with near horizon geometry $AdS_2\times S^2$.  
The most general field configuration, consistent
with the $SO(2,1)\times SO(3)$ symmetry of $AdS_2\times S^2$, 
is of the form:
\ben \label{e1}
&& ds^2\equiv g_{\mu\nu}dx^\mu dx^\nu = v_1\left(-r^2 dt^2+{dr^2\over 
r^2}\right)  +
v_2 \left(d\theta^2+\sin^2\theta d\phi^2\right) \nonumber \\
&& \phi_s =u_s \nonumber \\
&& F^{(i)}_{rt} = e_i, \qquad  F^{(i)}_{\theta\phi} = {p_i \over 4\pi} \,  
\sin\theta\, , 
\een
where $v_1$, $v_2$, $\{u_s\}$, $\{e_i\}$ and $\{p_i\}$ are constants.
For this background the nonvanishing components of the Riemann tensor 
are:\footnote{In our convention $R^\mu_{~\nu\rho\sigma}=\p_\rho
\Gamma^\mu_{\nu\sigma} - \p_\sigma\Gamma^\mu_{\nu\rho}
+\Gamma^\mu_{\tau\rho}\Gamma^\tau_{\nu\sigma}-\Gamma^\mu_{\tau\sigma}
\Gamma^\tau_{\nu\rho}$ where $\Gamma^\mu_{\nu\rho}$ is the
Christoffel symbol.}
\ben \label{e1a}
R_{\alpha\beta\gamma\delta} &= &-v_1^{-1} (g_{\alpha\gamma} g_{\beta\delta} 
- g_{\alpha\delta} g_{\beta\gamma})\, , \qquad \alpha, \beta, \gamma, 
\delta =r, t\, ,
\nonumber \\
R_{mnpq} &=& v_2^{-1}\,  (g_{mp} g_{nq} - g_{mq} g_{np} )\, , \qquad m,n,p,q
= \theta, \phi \, .
\een
It follows from the general form of the background that the covariant 
derivatives of 
the scalar fields $\phi_s$, the gauge field strengths 
$F^{(i)}_{\mu\nu}$ and the Riemann tensor
$R_{\mu\nu\rho\sigma}$ all vanish for the near horizon geometry. By the general
symmetry consideration 
it follows that the contribution to the equation of motion from
any term in the action that involves covariant derivatives of the gauge field
strengths, scalars
or the Riemann tensor vanish identically 
for this background and we can restrict our
attention to only those terms which do not involve 
covariant derivatives of these fields.\footnote{We are
assuming that all terms in the 
action depend
explicitly only on the gauge field strengths and not on gauge fields. This 
condition is
violated for example in string theory
by Chern-Simons type coupling of the gauge fields to three form field
strengths. However, as is well known, 
we can get rid of such terms by dualizing the two form field to a scalar
axion $a$. This field couples to the gauge fields only through field strengths.
If we encounter a theory where it is impossible to carry this out for all 
fields, our
analysis will still be valid if these additional terms do not affect the 
equation of motion and the entropy
for the specific black hole solution under study.} 

Let us
denote by $f(\vec u, \vec v, \vec e, \vec p)$ the Lagrangian density 
$\sqrt{-\det g}\,
\LL$ evaluated for
the near horizon geometry \refb{e1} and integrated over the angular 
coordinates\cite{0505122}:
\be \label{e2}
f(\vec u, \vec v, \vec e, \vec p) = \int d\theta\, d\phi\, \sqrt{-\det 
g}\, \LL\, .
\ee
The scalar and the metric field equations in the near horizon geometry 
correspond to
extremizing $f$ with respect to the variables $\vec u$ and $\vec v$:
\be \label{e3}
{\p f\over \p u_s} =0, \qquad {\p f\over \p v_i} = 0\, .
\ee
On the other hand the  non-trivial components of the gauge field equations
and the Bianchi
identities take the form:
\be \label{e4}
\p_r \left({\p \sqrt{-\det g} \, \LL\over \p F^{(i)}_{rt}}\right) = 0, \qquad 
\p_r F^{(i)}_{\theta\phi} = 0\ .
\ee
Both sets of equations  in \refb{e4} are automatically
satisfied by the background \refb{e1}, with the constants of integration having
the interpretation  as electric and magnetic charges of the black hole. 
{}From this it follows 
that the constants $p_i$ appearing
in \refb{e1} correspond to magnetic charges of the black hole, and
\be \label{e5}
{\p f\over \p e_i} = q_i\, 
\ee
where  $q_i$ denote the electric
charges carried by the black hole.

For fixed $\vec p$ and $\vec q$,
 \refb{e3} and \refb{e5} give a set of equations which are equal in number
to the number of 
unknowns $\vec u$, $\vec v$ and $\vec e$. In a generic case we may
be able to solve these equations completely to determine the background in 
terms of
only the electric and the magnetic charges $\vec q$ and $\vec p$. 
\footnote{We should
note however that the situation in string theory is not completely 
generic. For example
in $\NN=2$ supersymmetric string theories there is no coupling of the 
hypermultiplet
scalars to the vector multiplet fields or the curvature tensor to lowest 
order in $\alpha'$,
and hence in this 
approximation the function $f$ does not depend on the hypermultiplet
scalars. Thus the equations \refb{e3}, \refb{e5}
 do not fix the values of the hypermultiplet scalars in this approximation.}
This is consistent
with the attractor mechanism for supersymmetric background which says that the
near horizon configuration of 
a black hole depends only on the electric and magnetic
charges carried by the black hole and not on the asymptotic values of 
these scalar
fields. We shall return to a more detailed discussion of this mechanism 
in section \ref{s3}.

Let us now turn to the analysis of the 
entropy associated with this black hole. A general
formula for the entropy in the presence of higher derivative terms has 
been given in
\cite{9307038,9312023,9403028,9502009}. The formula simplifies enormously 
here since
the covariant derivatives of all the tensors vanish, and we get a simple 
formula:
\be \label{e6}
S_{BH} = 8\pi\, {\p \LL\over \p R_{rtrt}} \, g_{rr} \, g_{tt} \, A_H\, ,
\ee
where $A_H$ is the area of the event horizon and
${\p\LL\over \p R_{\mu\nu\rho\sigma}}$
is defined through the equation
\be \label{ex1}
\delta\, \LL = {\p\LL\over \p R_{\mu\nu\rho\sigma}} \, \delta\, 
R_{\mu\nu\rho\sigma}\, .
\ee
In computing $\delta\LL$
we can ignore all terms in $\LL$ which involve covariant derivatives of the
Riemann tensor, and treat the components of the Riemann tensor as independent
variables.

In order to simplify this
formula let us denote by $f_\lambda(\vec u,\vec v,\vec e, \vec p)$ 
an expression similar to the right hand side of \refb{e2} except 
that each factor of
$R_{rtrt}$ in the expression of $\LL$ is multiplied by a factor of 
$\lambda$. Then we
have the relation:
\be \label{e7}
\left . {\p f_\lambda
(\vec u,\vec v,\vec e, \vec p) \over \p\lambda} \right|_{\lambda=1}
=  \int d\theta \, d\phi\, \sqrt{-\det g} 
\, R_{\alpha\beta\gamma\delta} \, {\p \LL\over
\p R_{\alpha\beta\gamma\delta} }\, ,
\ee
where the repeated indices $\alpha,\beta,\gamma,\delta$ are summed over 
the coordinates $r$ and $t$.  Now since by symmetry consideration
$(\p \LL / \p R_{\alpha\beta\gamma\delta})$
is proportional to $(g^{\alpha\gamma} g^{\beta\delta} - g^{\alpha\delta} 
g^{\beta\gamma})$, we have
\be \label{e8}
{\p \LL \over \p R_{\alpha\beta\gamma\delta} }= - v_1^2 \, 
(g^{\alpha\gamma} g^{\beta\delta} - g^{\alpha\delta} 
g^{\beta\gamma}) \, {\p \LL \over \p R_{rtrt}}\, .
\ee 
The constant of proportionality has been fixed by 
taking $(\alpha\beta\gamma\delta)=(rtrt)$.
Using \refb{e1a}  and \refb{e8} we can  rewrite \refb{e7} as
\be \label{e9}
{\p \LL \over \p R_{rtrt}}\, A_{H} = {1\over 4} \, v_1^{-2} \, \left . {\p 
f_\lambda(\vec u,\vec v,\vec e, \vec p) \over \p\lambda} 
\right|_{\lambda=1}\, .
\ee
Substituting this into \refb{e6} gives\cite{0505122}
\be \label{e10}
S_{BH} = -2\pi\, \left . {\p 
f_\lambda(\vec u,\vec v,\vec e, \vec p) \over \p\lambda} 
\right|_{\lambda=1}\, .
\ee

We shall now reexpress
the right hand side of \refb{e10} in terms of derivatives of
$f$ with respect to the
variables $\vec u$, $\vec v$, $\vec e$ and $\vec p$. Since the
expression for $\LL$ is invariant under reparametrization
of the $r,t$ coordinates, every factor of
$R_{rtrt}$ in the expression for $f_\lambda$ must appear in
the combination  $\lambda\, g^{rr} g^{tt} R_{rtrt} = 
-\lambda v_1^{-1}$, every factor of $F^{(i)}_{rt}$ must appear
in the combination $\sqrt{-
g^{rr} g^{tt}}F^{(i)}_{rt}=e_i v_1^{-1}$, and every factor of
$F^{(i)}_{\theta\phi}=e_i$ and $\phi_s=u_s$ must appear without
any accompanying power of $v_1$. 
The contribution from all terms which
involve covariant derivatives of $F^{(i)}_{\mu\nu}$,
$R_{\mu\nu\rho\sigma}$ or $\phi_s$ vanish; hence there is no further factor of
$v_1$ coming from contraction of the metric with these derivative operators.
The only other $v_1$ dependence of $f_\lambda(\vec u,\vec v,\vec e, \vec p)$ is
through the overall multiplicative factor of $\sqrt{-\det g}\propto v_1$.  Thus
$f_\lambda(\vec u,\vec v,\vec e, \vec p)$ must be of the form $v_1 g(\vec u,
v_2,\vec p, \lambda v_1^{-1}, \vec e v_1^{-1})$ for some function $g$, and 
we have
\be \label{e11}
\lambda {\p f_\lambda(\vec u,\vec v,\vec e, \vec p) \over \p\lambda}  
+ v_1  {\p f_\lambda(\vec u,\vec v,\vec e, \vec p)  \over \p v_1}
+ e_i  {\p f_\lambda(\vec u,\vec v,\vec e, \vec p)   \over \p e_i}
- f_\lambda(\vec u,\vec v,\vec e, \vec p) = 0\, .
\ee
Setting $\lambda=1$ in \refb{e11}, using the equation of motion of $v_1$ 
as given in
\refb{e3}, and substituting the result into eq.\refb{e10} we get
\be \label{e12}
S_{BH} = 2\pi\, \left( e_i \, {\p f\over \p e_i} - f\right) \, .
\ee
This together with \refb{e5} shows
that $S_{BH}(\vec q, \vec p)/2\pi$ may be regarded as the 
Legendre transform of the function $
f(\vec u,\vec v, \vec e, \vec p)$ with respect to the
variables 
$e_i$ after eliminating $\vec u$ and 
$\vec v$ through their equations of motion \refb{e3}.

 The analysis can be easily generalized to higher dimensional theories as 
follows. In
 $D$ space-time dimensions we consider an extremal black hole solution 
with near
 horizon geometry $AdS_2\times S^{D-2}$. The relevant fields which can 
take non-trivial
 expectation value near the horizon are scalars $\{\phi_s\}$, metric 
$g_{\mu\nu}$, 
 gauge fields $A^{(i)}_\mu$ and $(D-3)$-form gauge fields 
 $B^{(a)}_{\mu_1\ldots \mu_{D-3}}$.
 If $H^{(a)}_{\mu_1
\ldots \mu_{D-2}}$ denote the field strength associated with the $B$ field,
 then the general background consistent with the $SO(2,1)\times SO(D-1)$ 
symmetry of
 the background geometry is of the form:
 \ben \label{e21}
&& ds^2\equiv 
g_{\mu\nu}dx^\mu dx^\nu = v_1\left(-r^2 dt^2+{dr^2\over r^2}\right)  +
v_2 \, d\Omega_{D-2}^2 \nonumber \\
&& \phi_s =u_s \nonumber \\
&& F^{(i)}_{rt} = e_i, \qquad  H^{(a)}_{l_1\cdots l_{D-2}} =
p_a  \, \epsilon_{l_1\cdots l_{D-2}}\, \sqrt{\det h^{(D-2)}}\,
/\Omega_{D-2}\,  .
\een
 where $d\Omega_{D-2}=h^{(D-2)}_{ll'} dx^l dx^{l'}$ 
 denotes the line element on the unit $(D-2)$-sphere,
  $\Omega_{D-2}$ denotes the area of the unit $(D-2)$-sphere,  
 $x^{l_i}$  with $2\le l_i\le (D-1)$
 are coordinates along this sphere and $\epsilon$ denotes
 the totally anti-symmetric symbol with $\eps_{2\ldots(D-1)}=1$. 
 We now define
 \be \label{e22}
f(\vec u, \vec v, \vec e, \vec p) = \int dx^2\cdots dx^{D-1}\, \sqrt{-\det 
g}\, \LL\, ,
\ee
as in \refb{e2}. 
Analysis identical to that for $D=4$ now tells us that the constants $p_a$
represent magnetic type charges carried by the black hole, and the 
equations which
determine the values of $\vec u$, $\vec v$ and $\vec e$ are
\be \label{e23}
{\p f\over \p u_s} =0, 
\qquad {\p f\over \p v_i} = 0\, , \qquad {\p f\over \p e_i}=q_i\, ,
\ee
where $q_i$ denote the electric charges carried by the black hole. 
Also using \refb{e6} which is valid in any dimension, we can show that
the entropy of the black 
hole is given by $2\pi$ times the Legendre transform of $f$:
\be \label{e24}
S_{BH} = 2\pi\, \left( e_i \, {\p f\over \p e_i} - f\right) \, .
\ee
as in \refb{e12}.

At string tree level, and in the absence of Ramond-Ramond
background fields (which includes all black holes
in heterotic string theory) 
the Lagrangian density at the horizon and hence the
function $f$ vanishes due to
the dilaton field equation. Thus eqs.\refb{e23}, \refb{e24} give:
\be \label{e35}
S_{BH} = 2\pi\, q_i \, e_i\, .
\ee
In other words the entropy of these black holes is given by $2\pi$ times the
product of the electric charge and the electric field at the horizon. It 
will be
interesting to see if this quantity admits a simple interpretation in the
world-sheet conformal
field theory that describes this background.

\sectiono{Attractor Mechanism and the Entropy Function} \label{s3}

We can now reformulate the attractor mechanism in a
more suggestive manner.
Let us define
\be \label{e31}
F(\vec u, \vec v, \vec q, \vec p) 
= 2\pi\, \left( e_i \, {\p f(\vec u, \vec v, \vec e, \vec p)
\over \p e_i} - f(
\vec u, \vec v, \vec e, \vec p) \right) \, ,
\ee
with $e_i$ determined by the equation:
\be \label{e32}
{\p f(\vec u, \vec v, \vec e, \vec p) \over \p e_i}=q_i\, .
\ee
In that case it follows from \refb{e23} that
the values of $\vec u$ and $\vec v$ at the horizon
are determined by extremizing
the function $F(\vec u, \vec v, \vec q, \vec p) $ with respect to $\vec u$
and $\vec v$:
\be \label{e33}
{\p F(\vec u, \vec v, \vec q, \vec p) \over \p u_s}=0\, , \qquad
{\p F(\vec u, \vec v, \vec q, \vec p) \over \p v_i}=0\, .
\ee
Furthermore, eq.\refb{e24} shows that
the black hole entropy $S_{BH}$ is given by the value of the function
$F$ at this extremum:
\be \label{e34}
S_{BH} (\vec q, \vec p)= F(\vec u, \vec v, \vec q, \vec p)\, ,
\ee
with $\vec u$, $\vec v$ given by  eq.\refb{e33}. This suggests that we call
$F(\vec u, \vec v, \vec q, \vec p)$ the entropy function.
Finally, the near horizon electric field $e_i$ are given by
\be \label{ei}
e_i = {1\over 2\pi} \, {\p F\over \p q_i}\, .
\ee

\sectiono{Relation to Earlier Results}

We are now in a position to discuss
the relation between our results and the observation of
\cite{0405146} that the Legendre transform of the entropy of a
black hole in $\NN=2$ supersymmetric string theory is given by the 
imaginary part of the generalized prepotential
of the theory. In the argument of the prepotential the real parts of the 
complex
vector multiplet scalar fields 
are replaced, up to a constant of proportionality, 
by the
magnetic charges of the black hole, whereas the imaginary parts of
these scalar fields
are replaced by the variables conjugate 
to the electric charges of the black hole. 
This
result follows from our results together with the following observations
(see {\it e.g.}
\cite{0007195}):
\begin{enumerate}
\item For the near horizon configuration of the black hole in $\NN=2$ 
supersymmetric 
string theory, all 
terms in the Lagrangian density vanish, except for a single term
proportional to the imaginary part of the generalized prepotential .
\item For the near horizon geometry the real parts of the vector multiplet
scalar fields are proportional to the  magnetic field at the horizon 
whereas the imaginary
parts of these scalar fields are proportional to the electric field at the 
horizon. 
\end{enumerate}
A little 
algebra shows that all the normalization factors also work out correctly and
we can reproduce the abovementioned observation
of \cite{0405146} from our results.

\medskip

{\bf Acknowledgement}: I wish to thank Rajesh Gopakumar for his comments
on the manuscript.
I also wish to thank the members of the
Center for Theoretical Physics at
MIT 
and DAMTP at Cambridge University 
for
discussion during various stages of this work.
The work was supported in part by the Jane Morningstar
visiting professorship at the Center for Theoretical Physics at MIT.

\end{document}